\documentclass[conference]{IEEEtran}
\IEEEoverridecommandlockouts
\pdfoutput=1
\usepackage{amsmath,amssymb,amsfonts}
\usepackage{graphicx}
\usepackage{adjustbox}

\makeatletter
\IEEEtriggercmd{\reset@font\normalfont\fontsize{7.7pt}{7.1pt}\selectfont}
\makeatother
\IEEEtriggeratref{1}

\begin{document}

\title{A Taxonomy for Understanding the Security Technical Debts in Blockchain-Based Systems\\
}

\author{\IEEEauthorblockN{Sabreen Ahmadjee}
\IEEEauthorblockA{\textit{School of Computer Science} \\
\textit{University of Birmingham}\\
Birmingham, UK \\
sxa1002@cs.bham.ac.uk}
\and
\IEEEauthorblockN{Rami Bahsoon}
\IEEEauthorblockA{\textit{School of Computer Science} \\
\textit{University of Birmingham}\\
Birmingham, UK \\
r.bahsoon@cs.bham.ac.uk}
}
\maketitle

\begin{abstract}
Blockchain is a disruptive technology intended at implementing secure decentralized distributed systems, in which transactional data can be shared, stored and verified by participants of a system using cryptographic and consensus mechanisms, elevating the need for a central authentication/verification authority. Contrary to the belief, blockchain-based systems are not inherently secure by design; it is crucial for security software engineers to be aware of the various blockchain specific architectural design decisions and choices and their consequences on the dependability of the software system. We argue that sub-optimal and ill-informed design decisions and choices of blockchain components and their configurations including smart contracts, key management, cryptographic and consensus mechanisms, on-chain vs. off chain storage choices can introduce security technical debt into the system. The technical debt metaphor can serve as a powerful tool for early, preventive and transparent evaluation of the security design of blockchain-based systems by making the potential security technical debt visible to security software engineers. We review the core architectural components of blockchain-based systems and we show how the ill-choice or sub-optimal design decisions and configuration of these components can manifest into security technical debt. We contribute to a taxonomy that classifies the blockchain specific design decisions and choices and we describe their connection to potential debts. The taxonomy can help architects of this category of systems avoid potential security risks by visualising the security technical debts and raising its visibility. We use examples from two case studies to discuss the taxonomy and its application. 
\end{abstract}


\section{Introduction}
Blockchain (BC) technology has been receiving wide academic and industrial recognition ever since the success of Bitcoin (a digital cryptocurrency), a seminal application based on BC technology.  The use of BC technology has gone beyond cryptocurrency systems to underlie many mainstream dependable software systems including finance, education, healthcare, transport, homeland security, identity management, etc. Organisation has been leveraging BC and its built-in capabilities, as a critical component within the software system architecture, to provide more dependable/secure computation and storage. The use of BC in an application can be “computationally active” e.g. the use of blockchain for active verification -- e.g. proof-of-work or proof-of-stake \cite{Xie2018} or “computationally passive” serving as immutable storage for critical and sensitive data \cite{Xu2017}. 

Although BC technology has distinct and inherent features that promise to produce systems that are cryptographically secure, ill-informed, suboptimal or wrong design decisions that relate to the choice, usage, configurations of a blockchain and its components are arguably the root cause of security technical debt, explicating this category of system. Blockchain components can, for example, include smart contracts, key management, cryptographic and consensus mechanisms, on-chain vs. off chain storage choices. 

Contrast to the belief, BC is not inherently secure \cite{International2017}. Evaluating the security of the design decisions for BC-based systems is crucial step for realising security requirements. As an example, verification design decisions within blockchain can take the form of proof-of-work or proof-of-stake \cite{Xie2018}. The verification process can rely on the wisdom and trust of peers. It can suffer from unnecessary redundancy, wrong assumptions and suboptimal design decisions of trust and verification process. Another complication is that these design decisions need to be weighed against those of performance, security, cost, energy efficiency among the many quality related concerns. It is imperative that exhaustive verification can be expensive to achieve, and designers often take simplified assumptions and decisions that vary in their level of trust and rigorous verification. It is also obvious that the architects and designers need to consider other quality related trade-offs that can compromise the dependability of the BC-based systems. Henceforth, we view suboptimal/ill-informed design decisions incorrect/imperfect assumptions and unmanaged trade-offs for as forms of security design debts that can subsequently introduce potential security risks into the system. Adversaries can for example, exploit the limitations in the envisioned design and verification to compromise the security of the system. 

The novel contribution of the paper is a taxonomy that defines and classifies the major properties and components of BC-based systems and describes how suboptimal and ill-design decisions, choices and imperfect assumptions can manifest into (security) technical debts for this category of systems. 

Software architects and designers for BC-based systems can leverage the technical debt metaphor to understand and evaluate the consequences of their design choices and decisions on the security posture and the fundamental objective of adopting blockchain to induce a system. The taxonomy can assist designers and architects to take early interventions, starting from the inception and design stages through making the potential security technical debt visible to security software engineers and software system’s adopters. Security Software Engineers can then better assess the security implications of the design decisions of building BC- based systems and various dependability trade-offs.  The taxonomy is indeed a novel attempt to rethink security technical debt in the context of blockchain by examining the unique characteristics of this disruptive technology and the way they induce the design decisions of this category of systems and investigating how these might incur security TD. Our taxonomy should help the architect of a BC-based systems to evaluate their design choices and decisions by investigating the potential presence of security TD. It can also help in understanding ways the debts can be accumulated and their ramifications on the security of the system. Our investigation reveals that BC-based systems can be more susceptible to security debt by design because this category of systems can inherently carry security technical debts that resembles of traditional secure systems, plus an additional set of blockchain specifications. 

Different from previous works on security technical debt \cite{nord2016} \cite{izurieta2018}, the paper is the first to discuss technical debt in the context of blockchain architecture. This paper follows a conceptual-to-empirical approach: it  first describes security TD, its causes, sources, and management; it then presents a taxonomy of security design decision of BC-based systems and argues how these decisions may manifest into debt. The taxonomy was developed following a review of publicly available knowledge on BC, including academic and industrial literature. We use scenarios and examples from two cases studies to discuss the taxonomy and illustrate its use.  

\section{SECURITY TECHNICAL DEBT}
Investigations that relate to security TD are recent. Example of studies include \cite{nord2016}\cite{izurieta2018}. In this section, we reflect on earlier studies and discuss security TD as a prerequisite for understanding Blockchain security TD. We posit that BC-based systems TD is essentially a form of security technical debt. This debt should be studied in the context of design decisions, which are specific to blockchain, their properties and configurations. 

Security TD can be attributed to innocent design oversights, sub-optimal or imperfect design decisions which could at some point lead to security vulnerabilities. Incurring security TD can compromise security attributes, such as confidentiality, integrity, non-repudiation, accountability, and authenticity \cite{organisation2011}. Any accumulated amount of security TD will have to be repaid with interest in the future. Interest represents the undesirable effects of the extra effort/cost that must be exerted to patch up the system security flaws \cite{Besker2017}. Interest of security technical debt can span many dimensions including costly repairs to stabilize the security of the system (i.e. security maintenance costs), cost of rebuilding the trust and retaining what is lost from users, credibility and reputation damage (i.e, brand damage costs, follow up costs that relates to preventing misuse of leaked information).

Ineffective prioritization and mismanagement for security and associated quality trade-offs are among the major causes of security TD. When security managers and architects have to prioritize tasks, they often give security implementation a lower priority than that of system functionalities. Software engineers often take security adjustment decisions under tight resources, time, and budget constraints. For example, managers may ignore employing a more expensive but efficient countermeasure such as a circuit gateways firewall because they are uncertain of the likelihoods of future attacks. These savings and ill-informed decisions can potentially manifest into security TD. Furthermore, The violation of cybersecurity compliance rules, best practices, and ignoring security standards could also manifest into potential security TD. Lack of expertise or inadequate knowledge regarding the implementation of secure systems is another reason that security TD might accumulate \cite{Tom2013}. Additionally, design oversights and wrong assumptions about the domain and/or environment can manifest into security TD.  

One source of security TD in design could relate to the inappropriate choice of security design patterns. A chosen pattern, for example, might not be well suited for addressing the security requirements of the application domain or it may have been implemented incorrectly, exposing the system into security risks. Furthermore, the inadequate, deferred, or lack of testing of security attributes can also be a significant source of security TD \cite{Tom2013}.  

Security technical debt can be either intentional or unintentional \cite{Klinger2011}. Intentional can be in the advent, where the likely risks of not deploying the security countermeasures can be marginal or negligible and the debt and its consequences can be kept under control. Unintentional security TD can be accidentally introduced due to the ill-practices described in the previous section. This kind of TD is worse because it is not visible. If it remains invisible, the debt can be accumulated and may manifest into a significant security risk. Both intentional and unintentional security debts should be carefully managed; otherwise, a software can become vulnerable to attacks and the risk of interest accumulation would take various forms of damage - spanning technical, reputation, juridical, harm etc.   

\section{SECURITY TECHNICAL DEBT IN BLOCKCHAIN-BASED SYSTEMS}
\subsection{Blockchain Background}
Blockchain is a chain of ordered blocks, each block is connected to the previous block via a cryptographic hash of its content \cite{Xu2017}. Generally, each block contains a list of transactions, a hash of the current block, a hash of the previous block, a timestamp, and other information such as a nonce value \cite{zheng2017}. Each node participating in the blockchain network can create a transaction and then exchange it with other peers in the network. Since the BC technology is based on a peer-to-peer decentralized network, it does not depend on a central controller. All authorized nodes in the network need to reach a consensus, a mechanism to ensure that nodes are in agreement concerning the latest block to be linked in the chain \cite{ManavGupta2017}.

\subsection{Blockchain Security Debt}
There are several design properties that distinguish a BC-based system, such as decentralization, immutability, transparency, etc. \cite{zheng2017}. As blockchain technology has a complex infrastructure and multiple configurations, building systems based on blockchain requires careful evaluation for design decisions and choices to avoid potential security design debts, that can lead to potential security risks. \textit{Blockchain design security debt can be attributed to suboptimal or ill-informed design decisions regarding the design and/or choice of cryptographic algorithms, consensus mechanisms, issues and wrong assumptions related to the configuration of the chains, smart contract etc. If incurred, the debt can compromise the security posture of the system and/or negatively impact the system’s security attributes and trade-offs}. Security design decisions for BC-based systems can leverage our understanding to security TD analysis to make the potential security TD visible and manageable.    

\subsection{Taxonomy}
As Ernst et al. \cite{Ernst2015} emphasize, the greatest source of TD is the architectural decisions, due to the fact that these are difficult to resolve and can take many years to evolve. To minimize the accumulation and the impact of security TD incurred due to suboptimal architectural decisions when building BC-based systems, we propose a blockchain design taxonomy and discuss its related security TD. This taxonomy can help an architect to understand a blockchain technology by classifying its features, dimensions and configurations, and comparing them according to their security strengths and weaknesses, thus making the architect aware of potential security TD when architectural decisions are made. As it can be difficult to anticipate the attack landscape and the mind-set of the adversary, this taxonomy helps introduce suggestions to mitigate the potential security TD in relation to likely threats. A discussion of architectural design security TD for various BC-based systems is presented in the following sub-sections.  

\subsubsection{Blockchain Types}
Among the crucial design decision for blockchain based system is the choice of blockchain type. There are three basic types of blockchain technologies.

\textbf {Public Blockchain:} in this type, also known as a permissionless blockchain, any node can read, send, and verify transactions. Nodes are also able to participate in the consensus process. A public blockchain is fully decentralized; there is no central authority that controls the system (for example, by managing the membership privileges) \cite{zheng2017}.
This type is highly available since they have no signal point of failure; they can better survive in the advent of distributed denial of service (DDoS) attack. Additionally, public blockchains are more resilient against ransomware attacks; since the records are securely stored in every node in the blockchain network, it is difficult to lock down these redundant records across the whole network. Another security strength of public blockchains is their immutability \cite{ManavGupta2017}. Once the information is written in the block, it cannot be altered because it is stored in different nodes in the decentralized network. Moreover, modifying data in a single block requires modification of all blocks in the chain, since they are connected with a cryptographic hash function. Importantly, the level of integrity of information is high in public blockchains, as any node can verify that the data have not been tampered with. Since the records and their updates are available to the public, the level of transparency of this type of blockchain is high as well. 

However, the decision to consider public blockchain should be approached with caution. The architect may accelerate the selection without careful analysis of the security implications of the said choice, leading to security debt. For instance, in a public blockchain, the validators can be unknown; as a result, if the majority of the nodes in the network are malicious, they can dominate the system and control the consensus process. This is known as a majority attack or a 51\% attack \cite{Li2017}. Consequently, the attacker can perform a double-spending attack and stop miners, who validate transactions and propose new blocks, from mining any available blocks. Moreover, public blockchains can be vulnerable to Sybil attacks  \cite{English2018}, where the attacker creates a large number of pseudonymous identities which appear to be different nodes, while in reality they are all under the control of a single party. Therefore, the attacker can gain influence and control a majority of the nodes in the network.  Importantly, once public blockchain is selected, the selection should take into consideration likely attacks. Careful and managed selection shall provide built-in mechanism for defending against potential attacks. Henceforth, architects should evaluate for these extreme scenarios and trade-offs, when selecting a blockchain type. A misinformed selection that fails to defend against likely attacks or taking a shortcut by not implementing adequate defence measures against likely attacks can carry debt. 

Additionally, the decision for adopting this type of blockchain shall also consider the trade-offs between privacy and transparency that is required for a given context; a deviation from this requirement may enter the system into a debt. While public blockchains are highly transparent and the chained information is visible to other peers, privacy can be difficult to be achieved; extra mechanisms might be required to strike a right balance between the two. Consequences of taking the security TD may lead to costly repays in the form of reparation of (financial) loss, if double spending attack would; for example, take place, or to recover the lost data and reputation damage, if sensitive data would be leaked.

\textbf {Private Blockchain:} in this type, also knowns as permissioned blockchains, only one organization has writing permissions, while reading permissions can be public or restricted to a preselected set of readers. Since private blockchains are controlled by a single group, they are known as fully centralized blockchains  \cite{Wust}. These specific features of the private blockchains give them some security advantages. Since all validators are known- as they are all member of a single group- some types of attack are difficult to launch, such as the 51\% attack and selfish mining attack. Sybil attack is also easy to prevent, since a centralized authority controls the system and can verify each node, allowing or rejecting requests to join the network. Moreover, private blockchains provides a greater level of privacy, especially when read permission are also restricted  \cite{zheng2017}. However, the potential security debt can be attributed to the wrong trust assumptions when selecting participants in the network. Moreover,  if the number of nodes in the network is low, private blockchains can be more prone to certain attacks, such DDoS or ransomware attacks. As public verifiability is not required, the integrity of the system can only be ensured if the system is not compromised. 

There is another type of permissioned blockchain knowns as consortium blockchain \cite{ManavGupta2017}. This type of BC is partially centralized, meaning that rather than the system being controlled by a single organization, a group of pre-selected nodes from multiple organizations is responsible for consensus and block validation. Read and write permissions can be determined by the consortium; they can be public or limited to selected nodes in the network. Since it is partially decentralized, it provides better availability than private BCs. It also has better privacy than public BCs as it is partially private. By knowing the capabilities, security strengths and weaknesses of each type of BC technology, architects can choose the type that best matches the requirements of the BC-based applications they are attempting to build.  The architects may treat potential mismatches as a form of debt and use the mismatch information as a way to evaluate the consequences of taking the debts. The information can help architects to be aware of security TD that might be incurred as a result of their design choices.

\subsubsection{Storage and Computation}
Although blockchains have unique features such as transparency and immutability, there may also be limitations to their storage space and computational power. Security, data integrity and users’ privacy might be negatively impacted by this limitation and the decisions to store sensitive information on or off-chain. Henceforth, architects can evaluate the off- or on-chain trade-offs decisions. Mismanagement of this trade-offs can come with a security technical debt; the interest and cost of this debt can crosscuts dimensions that relate data integrity, governance, privacy etc.   

\textbf {Data Storage:} data can be fully stored on a blockchain. There are various ways to do this. In the case of Bitcoin, limited information can be embedded in a transaction. For instance, by using OP\_RETURN \footnote{https://bitcoinfoundation.org/core-development-update-5/}, which is a script that is used to sign a transaction output as invalid, it is possible to embed arbitrary data in a transaction. This script can be used to store up to 80 bytes, and that was changed to be 40 bytes. The cost of storing 80 bytes of arbitrary data is approximately \$0,036. However, if a large amount of data needed to be stored on-chain, storing them would become too expensive. For instance, storing the ownership of a digital artwork on a blockchain would be very costly due to its size. Additionally, the confidentiality and privacy of the data stored on a public blockchain cannot guaranteed, as the content is visible to every node on the chain. In some BC-based applications, data should be visible to specific nodes and not to all the nodes in the network. In these cases, storing data off-chain would help overcome or mitigate such limitations. 

Commonly, when there is a decision to include off-chain storage, it will be used to store raw data, while meta-data and hashes of raw data will be stored on-chain  \cite{Xu2017}. The off-chain data storage can be a centralized repository such as a private cloud. However, the shortcut of using a centralized solution, because it is easy to configure and mange, can come with a debt in scenarios where centralization decisions can suffer from a single point of failure, in the advent of attacks e.g. DDoS. Although the intermediaries cannot alter the data as their hashes are stored on-chain, it could still fail at some point to deliver data to certain nodes as the attack propagates in the network. The potential penalty of this debt can be high; its future repay may entail costly reconfiguration for more reliable storage. Another option would be to use peer-to-peer decentralized file sharing platform such as IPFS \footnote{https://ipfs.io/} (InterPlanetary File System). In such cases, each node would need to run the blockchain, IPFS, and the middleware that coordinates between them. The original data is stored in IPFS, while hashes of this data are stored on-chain. Thus, the decentralized delivery of off-chain data is achieved. 

\textbf {Transaction Computation:} on-chain transactions are processed on the blockchain and considered to be valid only if they are propagated on the public records. Transaction execution, validation, and consensus mechanism increase the response time, and require communication and execution overheads. Moreover, mining processing- the process of assigning new blocks to the chain- is expensive because it typically involves a fee. in some BC-based applications, participant nodes need to prove a property of their confidential data without propagating it. In this case, it would be feasible to leverage off-chain computation. Computations can be outsourced to a third party to hide the information used during execution and then to generate the result. A proof of correct execution, such as Zero-knowledge proofs \cite{Kosba2016} can then be employed, hence, instead of performing the computation on-chain, the system merely verifies the proof of correct execution on-chain \cite{Eberhardt2017}. Thus, privacy would be greatly enhanced. Choosing to place data and computation on-chain or off-chain is a critical architectural decision that involves several trade-offs that relate to cost, privacy, security and integrity of information. Such decisions, if not managed properly can be viewed as debt with varying consequence, depending on the application domain.

\subsubsection{Consensus Protocol}
while a blockchain is decentralized technology which does not rely on a centralized authority to manage, authorize, and verify the transactions, a fault-tolerant consensus protocol is required to assure that all nodes agree on the new blocks that are appended to the blockchain. There are several different consensus mechanisms in use in existing BC technologies. When and how a transaction is verified and becomes immutable is depends on the consensus mechanism.

Proof-of-work (PoW) mechanism is used in public blockchain systems such as Bitcoin. In Bitcoin, new blocks are created using PoW, which is a puzzle that is simple and direct to verify, but very difficult and takes a long time to be solved. To generate a new block, the miner must solve a PoW puzzle using a large amount of computing power. The block will then be broadcasted to the other nodes in the network to verify it in consensus \cite{zheng2017}. PoW protects the blockchain network form Sybil attack since the total number of nodes is not important to solving the PoW puzzle; instead, it is a question of the total amount of computational power. While it is easy for an adversary to generate multiple node identities, it is difficult for them to generate a large amount of computational power.  

The decision to implement the PoW mechanism should be carefully evaluated and managed. The ill-informed decision can bear security debt since PoW properties can be exploited to initiate a number of attacks, if the block size is misconfigured. In PoW, the principle is that no miner should have more than half of the total processing power; otherwise, such a miner could effectively control the system by accomplishing attacks such as the 51\% attack, double spending, and the selfish mining attack. In selfish mining, malicious miners collude to increase their benefits by enforcing the honest miners to waste their power on creating blocks that eventually will not be linked to the chain. Meanwhile, the selfish miners can keep their mined blocks private, trying to maintain a private branch that is longer than the public branch. The selfish miners, can then reveal their branch and the honest miners switch to it; thus, the selfish miners win and take the reward, while the honest miners lose and waste their power \cite{Li2017}. The choice of the block size configuration and the number of block confirmations in a configuration can have potential impact on resisting/confining attacks. Hence, such decisions can bear a debt if they are not properly evaluated against these scenarios or they lag behind what is optimal for a given context/situation.  

Alternative consensuses mechanisms can be used such as Proof-of-Stake (PoS) to mitigate the potential security debts. In this mechanism, which is also used in public blockchain, a miner must prove their ownership of the cryptocurrency and pay a certain amount of it in order to mine blocks. This cryptocurrency will either be returned to the miner as a bonus, if the block is validated, or it will be fined \cite{Li2017}. Interestingly, PoS increases protection against malicious attacks: executing the attacks would be very expensive, as the attacker would need to own large amounts of money to perform the attack. Moreover, a miner who possesses a large stake most likely would not attack the system - for example, via double spending- since the attacker would eventually suffer from this attack because it would decrease the value of the cryptocurrency. To protect this mechanism, from possible attacks, the additional factor with combination of stake size should be considered. 

There are other consensus mechanisms that suitable for permissioned blockchains such as Practical Byzantine Fault Tolerance (PBFT) \cite{zheng2017}. In this mechanism, the identity of each miner must be known. Noticeably, each consensus mechanism has unique security properties which need to considered when architectural design decisions are made. Otherwise, if an unsuitable mechanism is selected, or the necessarily countermeasures are not considered, the ill-practice can carry security TD, and serious attacks can exploit weaknesses in the design.  

\subsubsection{Protocol Configuration}
\hfill\\
\indent\textbf{\emph{Block Size:}}
     refers to the maximum number of transactions aggregated within a block. The system's throughput is sensitive to the block size. Public blockchains currently can only process 3-20 transactions per second, which is very low compared to other payment services such as Visa, which can accomplish 2000 transactions per second. Consequently, some public blockchains, like Bitcoin, have raised their throughput by increasing the block size from 1MB to 8MB \cite{Decker2013}. However, deciding to increase the size should be carefully considered and needs to  be evaluated in relation to their consequences on security, data governance and performance and relative  to requirements of the application domain. Decisions that are not optimal for the said trade-offs can come with debts. Large blocks size can cause slower propagation speeds, which in turn result in raising the number of stale blocks \cite{GervaisETHZurich}; these are blocks that are not joined to the longest chain, because of a conflict or concurrency. Stale blocks do weaken the security of the blockchain because they cause chain forks that decrease the growth of the main chain. This can consequently increase the chances of an attacker in the network performing possible attacks such as double spending. Obviously, there is a trade-off between security and throughput in the BC-based systems; henceforth, the architect needs to be aware of this trade-offs and discuss the potential related debts,  if such trade-offs is not properly managed, when deciding the block size configuration. 
     
    \textbf{\emph{Number of Block Confirmations:}}
     commonly, in the BC-based systems, a transaction is confirmed after waiting for a specific number of blocks to have been created once the transaction has joined the blockchain. This strategy has been used to guarantee that a transaction is attached to the longest chain securely in order to prevent double spending. Another strategy for transaction confirmation is to add a checkpoint to the blockchain \cite{Xu2017}. The transaction is accepted once the checkpoint is valid; otherwise, if the fork chain starts before the checkpoint appears, it will be rejected by all nodes. An architect should choose the confirmation strategy that is most suitable and has the promise to provide the optimal security to the BC-based applications. Deciding on the required number of blocks for confirmation is a critical design decision, a wrong decision aimed at accelerating the confirmations can bear debts.

\subsubsection{Cryptographic Components}
cryptography is a key component of blockchain, which delivers system security properties such as integrity and non-repudiation. Two main cryptographic primitives that play an important role in blockchain are the cryptographic hash function and digital signature. The decision of selecting the cryptographic algorithms for each aforementioned primitive should be approached with caution since not considering the potential vulnerabilities of each of them, the decision can bear security debt.

\textbf{\emph{Cryptographic Hash Function:}}
 in BC-based systems, the hash function is used for many operations. One property of the hash function is that it is hard to invert; however, it is easy to verify. Therefore, any miner can easily verify the correctness of the block. Another property is that the smallest change to the input will result in an entirely different output. As data in each block is hashed, any modification or tampering with the data can be detected by all of the nodes in the network. Additionally, calculating a hash of the block requires a hash value of the previous block, which makes the blockchain tamper-resilience linked list. Collision resilience is an important property of the hash function, meaning it is computationally infeasible to obtain the same hash output from two different inputs. However, the hash function is vulnerable to collision attacks. Unawareness of such vulnerabilities when selected hash function and their implication on the security properties of the system, such as integrity and immutability, can enter the system into security design debts. Once a first single collision has been found, multiple collisions can be generalized and eventually it becomes feasible to compute them. Furthermore, quantum computing will make some attacks much easier  \cite{Giechaskiel2016}. 
 
\textbf{\emph{Digital Signature:}}
   uses asymmetric-key cryptography, in which each node should have a paired of private key and public key. The private key should be kept secret since it is used to sign the transaction, while the corresponding public key can be used by any node in the system to confirm the ownership of the signed transaction and to verify that the transaction has not been modified or tampered with. The public key is mathematically derived from the private key. Although there is a connection between the two keys, the private key cannot be identified using knowledge of the public key. The key generation algorithm of a digital signature scheme should have a good randomness source to generate different key pairs for different users. Otherwise, a weak randomness source could allow an attacker to recover a user’s private key and sign transactions. Mayer \cite{Mayer2016} found a vulnerability in the elliptic curve digital signature algorithm (ECDSA), which is used in the Bitcoin system. This algorithm does not generate enough randomness during the signature process, which allows the attacker to discover the user’s private key. Henceforth, the choice of a wrong digital signature algorithm for a given application can come with debts if the level of required randomness for a given domain is not met.
  Symmetric-key cryptography can also be used in blockchains if the data needs to be encrypted to preserve confidentiality. Oversights of implementing it when it is required would increase the potential security debts and observed the violation of  confidentiality. 

It is important to anticipate the impact of broken primitives, so that effective and appropriate countermeasures can be put in place to reduce the potential security debt. In \cite{Giechaskiel2016}, the authors demonstrate a threat model for broken primitive on Bitcoin, such as the SHA256 hash function and ECDSA. In addition, they suggest a migration strategy in case a cryptographic primitive is broken.  

\subsubsection{Key Management}
there are several alternative ways to manage and store private keys, each of which has its own security properties. Investigating them carefully and analysing their possible security threats is crucial, mismanagement decisions can bear security debt. In BC-based systems, users can use a piece of software, called a wallet, to store their private keys securely. Public keys and associated addresses, which are derived from public keys by using a hash function, can also be stored in the wallet. Keys can be stored in the device wallet directly on the specific file on the disk; thus, the user can have full control over his keys. However, private keys might still be under threat, as there are several attacks related to blockchains that have been crafted to steal secret keys, and not necessarily target the blockchain itself. For instance, specifically crafted malware, or man-in-the-browser attacks \cite{English2018} that involves the use of malware. If the device is compromised, the attacker can steal all the keys. To mitigate the potential security debts as a result of the decision of employing a single key, multi-signature can be used. This means that multiple secret keys are needed to generate the signatures. M signatures out of N private keys are required to sign any transaction. A simple example is that two-factor wallet that requires two devices-such as user’s mobile phone and laptop- to sign any transaction. However, this scheme increases the transaction’s size and negatively affects the confidentiality of the transaction, since it will be visible in the public block that a multi-signature transaction has been used. 

Alternative signature schemes such as threshold signatures \cite{Gennaro2016} can be used to minimize debts. In this scheme, a transaction can be signed using shares of a single private key. These shares are split among N parties using threshold cryptography. This scheme provides the same M-of-N security but increases the confidentiality since transactions look like normal transactions on the blockchain and the parameters M and N are kept private. Architects should provide a good enough security scheme to safeguard the private keys; otherwise, increasing rework cost of the unpaid TD would be accrued especially if the potential attacks are taken place, such as the attack that breached Bitfinex's cryptocurrency exchange \cite{International2017}. 120.000 bitcoins disappeared from users’ accounts because of this attack. Therefore, extra effort should be approached to resituate the loss and rebuild the users trust. 

\subsubsection{Smart Contract}
a smart contract is a piece of code and instructions that determines how and when data should move, without the need for a central authority to approve the instructions. The smart contract shows the blockchains’ participants whether or not specific conditions were met. If the conditions are not met, the smart contract has the capability to lock an asset \cite{Xie2018}. 
Each deployed smart contract should have a unique address to allow a user to interact with it through transactions. Smart contracts can be used in permissionless blockchains such as Ethereum \footnote{https://github.com/ethereum/wiki/wiki/Ethash} , which established its own blockchain with a built-in Turing-complete language to program its smart contracts. To execute the code of a smart contract, users should pay out a certain amount of cryptocurrency based on the underlying BC platform, such as the ETH cryptocurrency used in Ethereum. The execution time should be limited based on the code complexity. Once the limit has been exceeded, execution should stop. This mechanism prevents malicious code from being applied to smart contracts, such as consuming all resources by coding an infinite loop to apply DDoS attack. Smart contracts can also be used in permissioned blockchains such as Hyperledger Fabric \footnote{https://www.hyperledger.org/projects/fabric}, when the data of the contract need to be kept private and only a limited number of known nodes need to know about it.

Even though the smart contract is an important component in the second generation of blockchain technology, due to its powerful capabilities, it may still have several security vulnerabilities. Therefore, it is crucial to perform attack analyses for smart contract. The architect may use attack trees to visualise the likely attacks, their consequences and to design mitigation strategies. Twelve forms of vulnerability were systematically studied by Atzei et al. \cite{Atzei}. Moreover, 8833 Ethereum smart contracts were shown to be vulnerable, as Luu et al. \cite{Luu} discovered. They discovered several security bugs, which were the main cause of the security vulnerabilities. These bugs included timestamp dependence: since some smart contracts’ conditions based on a timestamp being triggered, if an adversary could alter it, the smart contract would become vulnerable. The lack of analysis may ship the system with vulnerabilities in the design, not providing good enough defence mechanisms against common attacks. This practice is an attribution to debts.

\begin{table*}[htbp] 
\caption{Evaluation of uPort Design Decisions and Choices }
\centering
\begin{adjustbox}{width=0.9\textwidth}
\small
\begin{tabular}{|l|l|l|l|l|} 

\hline
\textbf {\textit{BC Design Architecture}}& \textbf{\textit{Design Decisions and Choices}}& \textbf{\textit{Decisions Evaluation}}&
\textbf{\textit{Consequences}}&
\textbf{\textit{Examples of Debt}}\\
& &  \textbf{\textit{Using a Taxonomy}} & &\textbf{\textit{Mitigation}} \\

\hline
BC Type& Public permissionless, Ethereum  & Potential security TD is taken because of& Information disclosure. & Implementing symmetric   \\
    &   & the mismanagemen of the trade-off  &  Privacy and confidentiality  & encryption. \\
    & & between privacy and transparency by not & breach. &  \\
    & &  applying the secure mechanism to & & \\
    & &  preserve data privacy. See section III.C.1 & & \\
    
\hline
Data Storage & On-chain: Controller, Proxy, & Security TD is mitigated because of  & Increased system & - \\
             & and Registry smart contracts. & taking the right decision of employing   & dependability. & \\
            & Off-chain: IPFS       & the distributed off-chain storage. & &\\
            & &  See section III.C.2 & &\\
\hline
Consensus Protocol & PoW & Potential security TD might be taken & 	51\% attacks.  & Changing the consensus  \\
& & because of the misinformed decision & Breach of data integrity. & protocol.  \\
& & of the selected consensus mechanism.  & & \\
& & See section III.C.3 & & \\

\hline
Hash Function & SAH-3 & Potential security TD is taken because of & Collison attacks. & Implementing alternative  \\
 & & the misinformed decision of selecting & Breach of data integrity,  & secure hash function. \\
 & &  vulnerable choice. & immutability, and  & \\
 & & See section III.C.5 & non-repudiation.  & \\

\hline
Digital Signature & ECDSA & Potential security TD is taken because of& Recover sk. & Implementing secure  \\
& &  the misinformed decision of selecting& Breach system authenticity & digital signature.  \\
& &  vulnerable choice. See section III.C.5 &  and integrity. & \\

\hline
Symmetric Cryptography & Not used & Potential security TD is taken because of  & Information disclosure. & Implementing secure  \\
  &  & the accelerated decision of not applying  & Privacy and confidentiality  & symmetric encryption.\\
  & &  the secure mechanism to preserve data & breach. & \\
  & &  confidentiality. See section III.C.1 and III.C.5& &\\
 
\hline 
Key Management & pk, sk &  Potential security TD is taken & sk can be stolen. & Using multiple keys \\
&  & because of lag behind the optimal& Breach system authenticity& or key shares. \\
& & choice. Section III.C.6 & and integrity. &  \\
\hline
\end{tabular}
\end{adjustbox}
\label{tab1}
\end{table*}

\begin{table*}[t]
\caption{Evaluation of EMR System Design Decisions and Choices }
\centering
\begin{adjustbox}{width=0.9\textwidth}
\small

\begin{tabular}{|l|l|l|l|l|l}

\hline
\textbf {\textit{BC Design Architecture}}& \textbf{\textit{Design Decisions and Choices}}& \textbf{\textit{Decisions Evaluation}}&
\textbf{\textit{Consequences}}&
\textbf{\textit{Examples of Debt }}\\
& &  \textbf{\textit{Using a Taxonomy}} & & \textbf{\textit{Mitigation}}
\\

\hline
BC Type& Consortium permissioned, & Potential security TD might be taken because & Single point of failure. & Conducting a security  \\
 &  Hyperledger   &  of the suboptimal decision of selecting & 	DDoS attack. & analysis to decide  \\
 &   &  a limited number of participants. & Ransomware attack. & the right number of  \\
 & &  See section III.C.1 & Decrease system and/or   & participants on the  \\ 
 & & &data availability. & system. \\
 
\hline
Data Storage & On-chain: Logic and State & Potential security TD is taken because of & Single point of failure.& Using distributed storage.  \\
             &Chainecode  & the suboptimal decision of employing & DDoS attack. &  \\
             & Off-chain: local database, &  the centralized off-chain solutions. & Ransomware attack.& \\
             & Cloud.  & See section III.C.2  & Decrease system and/or &  \\ 
             & &  & data availability. & \\
\hline
Consensus Protocol & PBFT  & Security TD is mitigated because of the& Prevent Sybil attacks.  & - \\
& & chosen trusted validator (identity of  miners  & Increase systems integrity  & \\
& &  is known). See section III.C.2&and authenticity. & \\

\hline
Hash Function & AdHash/ MD5 & Potential security TD is taken because of  & 	Collison attacks. & Implementing alternative  \\
 & & the misinformed decision of selecting  & Breach of data integrity,  & secure hash function. \\
 & &  vulnerable choice (MD5).& immutability, and  & \\
 & & See section III.C.5 & non-repudiation.  & \\

\hline
Digital Signature & Unknown  & Taking a well-informed decision for choosing & -& - \\
& & the secure digital signature is crucial to avoid  & & \\
& &   security TD. See section III.C.5& & \\

\hline
Symmetric Cryptography & Used but the algorithm &Encrypting the stored data mitigates the & Increase data confidentiality & - \\
  & type is not specified & security TD. See section III.C.5 &and user privacy.  &  \\
 
\hline  
Key Management & Signature: \( pk_U^S \), \( sk_U^S \)& Potential security TD is taken because of & Secret keys could be stolen. &  Using multiple keys  \\
&  Asymmetric Encryption: \( pk_U^\varepsilon \), \( sk_U^\varepsilon \) & lag behind the optimal choice. & Breach system authenticity & or key shares. \\
& Symmetric Encryption: \( SK^ {A\varepsilon S} \)&  Section III.C.6 & and integrity.& \\

\hline
\end{tabular}
\end{adjustbox}
\label{tab2}
\end{table*}

\section{CASE STUDIES}
In the following section, we show how the taxonomy can help architects and designers to identify potential security debts in BC-based systems and mitigate the consequences of that debt. We use examples from two case studies to show how the taxonomy can be mapped to discuss core design decisions and choices related to BC components and configurations and their attribution to security debt.     

\subsection{uPort Architectural Design and Analysis}
uPort \footnote{https://www.uport.me/}is an open-source BC-based system for self-sovereign identity. It aims to return ownership of digital identities to individuals and eliminate the need for the centralized intermediaries \cite{Heck2016}. The public permissionless Ethereum BC is the underlying platform that was used to build uPort. Therefore, the system is available to anyone to create and store identities, which means all participants are unknown. Since the information is publicly available, all nodes can validate the state of the chain and any identity can be verified. As uPort was built on top of Ethereum, a PoW consensus mechanism was used, which provides security against Sybil attack, however, if one node or group of malicious nodes have more than 50\% of the hashing power, they can control the entire system, as PoW is prone to 51\% attack unless an appropriate countermeasure is already in place as discussed in section III.C.3. 

The data associated with the identity are stored off-chain on IPFS and the data can be retrieved by their hashes. In Ethereum, the Keccak \footnote{https://github.com/ethereum/wiki/wiki/Ethash} hash function is used, and to sign any transaction, the ECDSA scheme is used which, has certain vulnerability as mentioned previously. Any transaction occurring in Ethereum requires a payment of ETH cryptocurrency. Three smart contracts were designed in uPort: Controller, Proxy, and Registry. To generate a new user identity, the uPort mobile application generates the public and private key pair, then sends the transaction to the BC which initiates a new Controller to store a reference to the public key. After that, a new Proxy is generated which stores a reference to the Controller contract that was just generated. The Proxy address includes the unique uPort identifier (uPort-ID), whereby a user can create multiple of them. The user's private key is stored only on her mobile device. A key recovery protocol is in place in order to recover a user’s key in cases in which it has been lost or stolen. In such cases, the Controller should retain a list of recovery trustees, such as the user's friends, who can help the user to recover her keys. If a quorum is completed, the Controller can remove the lost public key and store a new one by invoking a dedicated function on the Proxy. Thus, a persistent uPort-ID can be preserved even after the loss of cryptographic keys.

However, in the uPort architecture, all the user's recovery trustees are publicly available in the Ethereum. This transparency allows the attacker to target a user’s trustees to obtain their identities. If an adversary can compromise a mobile application and change a user’s trustees through the Controller, the uPort-ID will be compromised permanently. The third smart contract is a Register which stores the hash of the identity attributes and maps them to users’ uPort-IDs. This map might leak information about attributes and their relationships to the identity provider.

\subsection{Electronic Medical Record Sharing Architectural Design and Analysis}
The second case study is a framework that manages and shares electronic medical record (EMR) data for cancer patient care. This framework was proposed by Dubovitskaya et al. \cite{Dubovitskaya} in collaboration with a Stony Brook University Hospital. They utilize BC technology to maintain immutable and verifiable records that keep track of all actions arising across the network. This helps to improve the integrity of distributed sensitive medical data, reducing the time needed to share EMRs and the overall cost. This medical application was built on top of the Hyperledger permissioned BC, in order to safeguard the privacy of the highly sensitive data. Additionally, a fast response time is highly necessary in the medical system, which can be provided easily by a permission BC. Also, in permissioned BCs, there is no need to pay for a transaction's execution and this helps increase the usability of that system.\\
\indent Since all users in a medical application are known (patients, doctors) and only a predefined set of nodes are allowed to participate in the PBEF consensus mechanism, security attacks such as 51\% and Sybil attacks are unlikely to occur as previously discussed. In the Hyperledger BC, smart contracts were developed in the form of Chaincode, which comprises of Logic and correlated State components. Logic is a set of instructions that allow the patient to specify fine-grained access control policy for his own data and permit efficient data sharing. The State of the Chaincode saves the information as key-value format, where the key is a patient ID in the system and the value is a patient’s record, and it also includes corresponding block number. When creating a new user, the Membership Service component of the medical application is responsible for registering the user according to his role.\\
\indent A registered doctor can upload, access, and share data based on the permissions defined by the patient. The Logic of the Chaincode does the verification of the access control rights. Each doctor has a public key \( pk_U^S \) and a private key \( sk_U^S \) for signing, as well as a \( pk_U^\varepsilon \) and \( sk_U^\varepsilon \) for encryption. Patient can generate a metadata record on the Chaincode, retrieve it, and specify permissions. In addition to the two aforementioned key pairs, patient also has a symmetric encryption key \( SK^ {A\varepsilon S} \) that encrypts and decrypts patient data. If a patient needs to enable a doctor to access his data, he should encrypt the \( SK^ {A\varepsilon S} \) with the encryption public key of the doctor \( pk_D^\varepsilon \), and then share the encrypted value with the doctor. Since each user has a single key pair for signing and another single key pair for encryption, these keys constitute a single point of failure. If an attacker success to compromise them, he could sign and encrypt data himself. Moreover, if a passive attacker compromises the symmetric key of the patient, he could decrypt and observe data.\\
\indent Patient metadata is stored on-chain, while two off-chain storage locations are used to store the patient’s raw data: an in- hospital database that stores the oncology-related data, and a cloud storage database that saves a patient’s data that and is organized according to specific categories of the data and encrypted with the patient’s symmetric key. A doctor can access data in the cloud storage according to the permission policy specified by the patient and implemented in the Chaincode Logic. These two off-chain storages reintroduce centralization into the BC-based system and can function as a single point of failure and prone to attacks such as DDoS attack.\\ \indent Table \ref{tab1} and Table \ref{tab2} demonstrate the evaluation of the design choices and decisions of the two case studies by using a taxonomy that has been presented in section III.C. Consequences of the security debts and examples on how to mitigate them are also illustrated in the tables. It is worth noting that other debt mitigation strategies can be possible, we only look at sample examples.  

\section{Discussion}
The consequence of the debt can crosscut several dimensions such as confidentiality, integrity, non-repudiation, and authenticity. The consequence can be discussed in relation to the core objective of the applications and the enhanced security by inducing the system with blockchain. For the simplicity, the three main security requirements, known as CIA, an acronym corresponding to confidentiality, integrity and availability, that any security design should address are considered.    

\textbf {Confidentiality:} effective security defences should be considered to prevent unauthorized users from reading sensitive data. Suboptimal decision due to the application of weak (not secure enough) security mechanism (e.g. symmetric encryption) can carry a security debt, where the debt symptoms can be observed on breaches to confidentiality. Since uPort was built on the public blockchain, any participating node with the chain can access the data stored in the Registry smart contract, where the overall data structure is visible. As discussed in section III.C.1 this design should be evaluated taking into account that node are not necessarily trusted. Henceforth, the design should provide built-in mechanism to safe guard the system against mistrusted nodes. The evaluator should also consider managing the trade offs between transparency and privacy as discussed in Section III.C.1. Inappropriate management for this trade-off can disclose the relationship between an identity provider and their metadata signaling a potential risk. In contrast, the EMR system was built on a permissioned blockchain, which preserves the confidentiality of the patient’s data by encrypting it before storing it and allowing the patient to define fine-grained access control over his data. However, if a doctor’s \( sk_D^\varepsilon \) is stolen, an attacker can decrypt the patient’s data, because the patient encrypts his symmetric key \( SK^ {A\varepsilon S} \)  by the doctor \( pk_U^\varepsilon \). Consequently, if the choice of the key is not analysed for potential risk, or by deploying the wrong key that does not guarantee security, the choice would bear a debt. This trial and error decision of using single key pair over the alternative available solutions such as multiple keys or key shares can exhibit high security debt in the design. Once the attack is accomplished, extra effort is required to provide a mechanism that recover the system from the loss of the confidential data and protect the users of whom their data may have been leaked.\\ \indent \textbf {Integrity:} taking suboptimal design decision of applying ineffective safeguarding mechanisms against potential data manipulation due to the time or budget limitations can introduce security debts into the systems.  Analysing the effectiveness of the chosen mechanism is necessary to avoid security technical debt. If the mechanism has proven to be ineffective from the security point of view, alternative choices should be considered to provide better integrity and minimising the security debt and potential consequences. In uPort any tampering with data in a block is visible to the whole nodes that participate in this public BC network. Also, any participant can verify the integrity of a distributed identity and the ownership of the signed transactions. These properties are achieved by using a consensus mechanism, hash function, and digital signature. As Ethereum is the underlying BC of uPort, the vulnerable ECDA scheme is used whereby an attack is be able to recover users’ private keys. The choice of this digital signature algorithm is a visible security technical debt as this signature is proven to be weak - see section III.C.5. Furthermore, the single private key is stored in the mobile application which can be compromised by attackers, allowing them to steal the private key and sign any transaction, as discussed section IV.A. Taking ill-informed or suboptimal decision of applying single key because it is easy to implement and manage would be a potential security debt on the system. Considering multi-signature is example of alternative solution that can mitigate the debts.\\ 
\indent Similarly, in the EMR system, a single key pair is used to sign a transaction, which makes it easier for the attacker to target a single location. For instance, if a doctor’s private key is stolen, the attacker could upload files and sign transactions as the doctor. Therefore, if the architect had not applied an adequate threat model to assess the security risk level of applying single key on the system and ill-informed or sub-optimal design decision had been taken, high security debt is carried. In case of an attack, an extra effort/cost becomes unavoidable to recover the system, retrieve original data, and protect the users of whom their data has been manipulated.\\
\indent \textbf {Availability:} taking well-informed decisions and applying mechanisms that ensure secure data and service availability is crucial for avoiding security technical debt. uPort’s availability is high because a fully decentralized and distributed public blockchain is used. Additionally, it uses distributed file sharing IPFS to store data off-chain. On the other hand, a local database and cloud storage are used to store data off-chain in the EMR system. This style of storage can suffer from a single point of failure, negatively affecting  the availability of the data and the service. For example, it is easier for the attacker to apply a DDoS attack for a single storage place to make the system unavailable. Therefore, if the availability impact of choosing this style of storage had not been carefully assessed, the system would incur security debt and an extra maintenance cost for recovering the system and remaintating availability level should be paid back. This is an addition to rebuild/retaining customer's trust. \\ 
\indent As it can be difficult to eliminate the security debt, visualizing the potential security technical debt of BC-based systems at early design stage is important. The above are just examples on how the taxonomy can be mapped to evaluate the key design decisions of BC-based system for debts and mitigate their ramifications through alternative design. 

\section{RELATED WORK}
Blockchain technology has been the focus of several researches. Two research studies \cite{Xu2017}\cite{Xu} represent a classification of the architectural properties of BC-based systems and show the impact of these properties on the performance and quality attributes of the system. Yet, their impact on the security properties and their consequential security risks has not been covered. In \cite{Li2017}, the authors reviewed the security threats to blockchain and the corresponding attacks, and suggested some security solutions for blockchain. Also, Lin et al. \cite{ManavGupta2017} have presented a general overview of blockchain and briefly explained some security attacks. However, none of these works address the problem of suboptimal design decisions of BC-based systems and their impact on different security requirements. This study has applied the TD metaphor to raise the visibility of possible security risks caused by the non-ideal design options of BC-based systems. Despite the vast contributions on technical debt in software, only a couple of studies has looked at security TD. In \cite{nord2016}, the authors applied technical debt knowledge to identify software vulnerabilities. Analysing technical debt correlated with security weaknesses had been investigated by \cite{izurieta2018}. Different from previous works on security technical debt, to our knowledge, we are the first to investigate security TD in BC-based systems architectures.  

\section{CONCLUSION}
We have argued that systems that leverage BC are not inherently secure by design. We have posited that the technical debt metaphor can serve as a powerful tool for evaluating the security design decisions and choices of blockchain-based systems. We have discussed how design decisions and choices of blockchain components and their configurations, covering smart contracts, key management, cryptographic and consensus mechanisms, on-chain vs. off chain storage choices can manifest into security technical debt. We have contributed to a taxonomy that classifies the blockchain specific design decisions and choices and their attribution to potential debts in the design. The taxonomy can help architects of this category of systems to avoid potential security risks by visualising the security technical debts and raising its visibility to ease TD management. We have used examples from two case studies to show how the taxonomy can be mapped to discuss core design decisions and choices related to BC type, data storage, consensus protocol, hash function, digital signature, key management and cryptographic choices and their attribution to security debt. We highlight the consequences of the debt and given examples on how to mitigate the debts.    

\let\secfnt\undefined
\newfont{\secfnt}{ptmb8t at 9pt}
\bibliographystyle{ieeetr}

\end{document}